\documentclass[]{article}
\usepackage{authblk}

\usepackage[english]{babel}
\usepackage{amsfonts}
\usepackage[letterpaper,top=2cm,bottom=2cm,left=3cm,right=3cm,marginparwidth=1.75cm]{geometry}
\usepackage{subcaption}
\usepackage{float}
\usepackage{xspace}
\usepackage[sort&compress,square,numbers]{natbib}
\usepackage{amsmath}
\usepackage{graphicx}
\usepackage[colorlinks=true, allcolors=blue]{hyperref}
\usepackage{booktabs}
\usepackage{mhchem}

\title{Enhancing \textit{ab initio} diffusion calculations in materials through Gaussian process regression}
\author[1]{Seyyedfaridoddin Fattahpour}
\author[1]{Sara Kadkhodaei}
\affil[1]{Department of Civil, Materials, and Environmental Engineering, University of Illinois Chicago, Chicago, IL, USA}
\date{}                     
\begin{document}
\maketitle

\begin{abstract}
Saddle point search schemes are widely used to identify the transition state of different processes, like chemical reactions, surface and bulk diffusion, surface adsorption, and many more. In solid-state materials with relatively large numbers of atoms, the minimum mode following schemes such as dimer are commonly used because they alleviate the calculation of the Hessian on the high-dimensional potential energy surface. Here, we show that the dimer search can be further accelerated by leveraging Gaussian process regression (GPR). The GPR serves as a surrogate model to feed the dimer with the required energy and force input. We test the GPR-accelerated dimer method for predicting the diffusion coefficient of vacancy-mediated self-diffusion in \textit{bcc} molybdenum and sulfur diffusion in hexagonal molybdenum disulfide. We use a multi-task learning approach that utilizes a shared covariance function between energy and force input, and we show that the multi-task learning significantly improves the performance of the GPR surrogate model compared to previously used learning approaches. Additionally, we demonstrate that a translation-hop sampling approach is necessary to avoid over-fitting the GPR surrogate model to the minimum-mode-following pathway and thus succeeding in locating the saddle point. We show that our method reduces the number of evaluations to a fraction of what a conventional dimer requires.  
%
%
\end{abstract}

\section{Introduction}

Transition state theory (TST) \cite{Eyring1935Activated} is widely used to quantify the free energy barrier (or activation free energy) of chemical reactions, such as molecular dissociation, as well as material processes including bulk diffusion, surface diffusion, or surface adsorption \cite{VINEYARD1957121,Jones2001Absolute,Kadkhodaei2018Simple}. Within TST,  the activated state is identified as the saddle point on the free energy surface. Consequently, saddle point search methods are crucial for quantifying the activated state, energy barrier, and rate of various kinetic processes in materials \cite{HANNESNudged,ren2013climbing,Henkelman2000climbing,Henkelman1999dimer,Caspersen2005Finding,Sheppard2008Optimization,Henkelman2000Improved}. Among saddle point search methods, minimum mode following methods \cite{Zeng2014Unification,Olsen2004Comparison}, such as the dimer algorithm\cite{Henkelman1999dimer,Xiao2014Solid-state-dimer,Plasencia2017Improved}, have gained popularity due to their computational advantages, particularly for solid-state processes. Unlike alternatives such as the partitioned rational function optimization (P-RFO), these methods do not require energy Hessian calculations in the high-dimensional space of solid-state atomic systems~\cite{Olsen2004Comparison}. However, utilizing the dimer algorithm can still be computationally prohibitive when combined with density functional theory (DFT) energy calculations. In this study, we show that we can further enhance computational efficiency by utilizing Gaussian process regression (GPR) as a surrogate model for inputting forces to the dimer algorithm. We implement a GPR-guided dimer algorithm, which we call the GPR-dimer, and apply it to investigate bulk diffusion in \textit{bcc} Mo as well as diffusion of sulfur in \textit{hexagonal} \ce{MoS2}. Building upon previous studies combining GPR with saddle point or minimum energy path search methods, this work provides two new insights for advancing the utility of GPR-dimer. Firstly, we employ a multi-task GPR learning approach, demonstrating a significant reduction in both training error and time compared to previously used GPR learning methods. Secondly, we introduce a translation-hop sampling approach that reduces the computational effort of DFT and enhances the robustness of the search algorithm. Furthermore, this work extends the application of GPR-dimer to solid-state materials.

Previous studies have successfully employed GPR to accelerate the search for saddle points or minimum energy paths~\cite{Koistinen2017NEB,Denzel2018Gaussian,Denzel2019Gaussian,Koistinen2019minimum_mode,Koistinen2019Nudged,Denzel2020Hessian}. For instance, Jónsson's group developed an adaptive GPR surrogate model of the potential energy surface (PES)~\cite{Koistinen2017NEB}. They utilized this model to derive an initial interpolation of the minimum energy path, which was subsequently optimized using the nudged elastic band (NEB) method. Their investigations focused on 25 chemical reactions, primarily involving organic molecules, known as the Baker test systems ~\cite{Baker1996location}. The results demonstrated the superiority of the GPR-accelerated NEB search over the classical NEB optimizer. The GPR model was trained using the Matérn covariance function and a predetermined weighted combination of energy-based and force-based loss functions. Another study by Kästner's group combined a GPR-interpolated PES with the P-RFO method to identify transition states in the Baker test systems~\cite{Denzel2018Gaussian}. By providing the necessary Hessian information to the P-RFO optimizer, the surrogate GPR model rendered the method computationally efficient, comparable to force-based methods like the dimer algorithm. Subsequently, they introduced a GPR-based Hessian update scheme~\cite{Denzel2020Hessian}, where the GPR was employed to update Hessian matrices using gradient-based information during the optimization procedure. This approach involved at least one initial Hessian, along with additional energies and gradients. Moreover, Denzel and Kästner~\cite{Denzel2018Gaussian} utilized the Matérn covariance function and the ``derivative observation'' GPR learning tehnique~\cite{SolakAdvances2002}, which explicitly relates the learned forces to the negative partial derivatives of learned energies. In a subsequent study, the same group combined the GPR surrogate model, employing derivative observation learning, with the NEB optimizer~\cite{Denzel2019Gaussian}. In two additional studies, Jónsson's group introduced the inverse-distance covariance function as an alternative to the previously employed covariance functions~\cite{Koistinen2019minimum_mode,Koistinen2019Nudged}, resulting in a significantly enhanced GPR surrogate model. By utilizing this improved GPR model to guide the dimer and NEB saddle point searches, they investigated the dissociative adsorption of an \ce{H2} molecule on the Cu(110) surface, three gas-phase chemical reactions, and the diffusion hop of an \ce{H2O} molecule on an ice Ih(0001) surface~\cite{Koistinen2019minimum_mode,Koistinen2019Nudged}. 
Building upon previous studies, we extend the application of GPR-dimer to investigate solid-state processes, moving beyond molecular processes. This study presents two examples of solid-state processes: vacancy-mediated self-diffusion in \textit{bcc} Mo and sulfur diffusion in \textit{hexagonal} \ce{MoS2}. We address the challenge of handling high dimensionality when applying GPR-dimer to solid-state processes through the use of the inverse-distance covariance function introduced by Jónsson's group \cite{Koistinen2019minimum_mode}. By only considering atoms in the vicinity of the diffusing atom, the inverse-distance covariance formulation significantly reduces the degrees of freedom in the high-dimensional space of the atomic systems considered in this study. More details are provided in section ~\ref{sec:method}. 
Additionally, this work advances the GPR-dimer method in two key aspects. First, we introduce the use of multi-task learning for the GPR surrogate model, resulting in a substantial improvement in the model's performance and robustness compared to the previously employed derivative observation learning. The multi-task GPR learning approach resembles the learning scheme used in Ref. \cite{Koistinen2017NEB}, where the loss function represents a weighted average of energy and force losses. However, in the multi-task approach, the contribution from force and energy losses is learned through a shared covariance function, unlike the approach in Ref. \cite{Koistinen2017NEB}, which requires prior knowledge of the contribution of each loss.
Second, we demonstrate that by employing a translation-hop sampling approach (defined below), the GPR-dimer search becomes both successful and robust. As detailed in section \ref{sec:method}, the GPR-dimer method iteratively updates the GPR model as the dimer walker progresses, incorporating new DFT-calculated values from the energy surface into the training data. We show that a minimum number of dimer translation steps must be hopped over before updating the GPR to ensure successful guidance of the dimer to reach the saddle point. We refer to this approach as the \textit{translation-hop sampling} approach. This sampling strategy strikes a balance between an overfitted and underfitted surrogate model. Sampling at every translation step leads to a GPR surrogate model that is overfitted to the dimer walk path on the PES, while skipping too many translation steps results in an underfitted model. A detailed discussion is provided in section \ref{sec:results}. Denzel and Kästner discuss a similar balance between interpolation and extrapolation with the use of an overshooting approach for sampling the GPR for geometry optimization \cite{Denzel2018geometry} (not for saddle point search).

The remainder of this article is organized as follows: In section \ref{sec:method}, we explain the GPR-dimer method developed in this study. In section \ref{sec:results}, we validate the predictions of our GPR-dimer method for diffusivity coefficient of monovacancy diffusion in bulk \textit{bcc} Mo and the activation energy for sulfur diffusion in \ce{MoS2}. Subsequently, in section \ref{sec:results}, we elucidate the role of different factors in enhancing the performance of the GPR-dimer saddle point search method. Finally, we compare the computational cost of the standard dimer method against our implementation of the GPR-dimer method. In section \ref{sec:discussion}, we provide a general interpretation of the numerical experiments using the GPR-dimer method within the context of GPR learning and the dimer search algorithm.
\section{Method}\label{sec:method}
The approach to accelerate the dimer walk using GPR is based on a simple premise: GPR serves as a surrogate model for the computationally intensive sampling of the potential energy surface typically through methods like DFT. Once trained, the surrogate model can readily provide the energy values and their gradients (forces) at unsampled locations of the energy surface. 
The interaction between GPR and the dimer takes place through an iterative feedback loop: GPR provides estimates of the energy and its gradient along the dimer walk, while the dimer walker contributes new points on the energy surface. These new points are sampled through DFT and then used to update (retrain) the GPR. Through this iterative process, the dimer gradually converges towards the saddle point. The next two subsections delve into the design and training of the GPR, as well as the communication between GPR and the dimer, respectively.
\subsection{Gaussian process regression (GPR) surrogate model}
\textbf{\textit{Training Data Set}}. The GPR is initially trained on $n_i$ atomic configurations. These atomic configurations are collected from the first $n_i$ translation steps of a DFT-guided dimer walk (i.e., standard dimer). The standard dimer is launched from an atomic configuration which is estimated to be in the vicinity of the saddle point using a geometric interpolation (as detailed in section~\ref{sec:results}). The atomic configurations constitute the input space and the DFT-calculated atomic forces and energies constitute the target values in the training data set (see more details below). The training data set is expanded as the dimer progresses by adding a new DFT-calculated data at every $n_h$ translation steps of the dimer walk. We call this approach the translation-hop sampling method. The effect of different $n_h$ values are examined and explained in section~\ref{sec:results}. For both studies of \textit{bcc} Mo self-diffusion and sulfur diffusion in \textit{hex} \ce{MoS2}, we use 3 atomic configurations for the initial training of the GRP ($n_i=3$) and we hop over 10 translation steps before adding a new DFT calculation to the training data ($n_h=10$).  

The DFT calculation of energy and forces are performed using the Vienna Ab-initio Simulation Package (VASP) \cite{kresse1996}, which employs the projector-augmented-wave (PAW) method \cite{blochl1994} and the generalized gradient approximation (GGA) for exchange-correlation energy in the Perdew-Burke-Ernzerhof (PBE) form \cite{perdew1996}. For \textit{bcc} Mo, we use a a $3\times3\times3$ supercell of the conventional bcc unit cell with 54 atoms. We use a Monkhorst-Pack \textit{k}-point mesh of 5 $\times$ 5 $\times$ 5 and an energy cutoff of 520, respectively, within the PBE exchange-correlation functional. For \textit{hex} \ce{MoS2},  we use a $2\times2\times2$ supercell of the convectional hexagonal unit cell with 48 atoms. We use a Monkhorst-Pack \textit{k}-point mesh of 5 $\times$ 5 $\times$ 1 and an energy cutoff of 520, respectively, within the PBE exchange-correlation functional.
\vspace{2mm}
\par
\noindent\textbf{\textit{GPR Covariance Function}}. Choosing an appropriate covariance function is crucial for GPR performance \cite{Rasmussen2006}. Here, we use the inverse distance covariance function of Ref.~\cite{Koistinen2019minimum_mode}, which demonstrates superior performance compared to the radial basis function (RBF) or its variants (e.g., Matérn) as shown in Ref.\cite{Koistinen2019Nudged}. Compared to a stationary covariance function such as RBF, the inverse distance covariance function can better capture the asymmetry of inter-atomic forces, specifically the large repulsive forces caused when atoms get close to each other. This is because the inverse distance difference measure (i.e., $\mathcal{D}_{1/r}\left(\mathbf{x}, \mathbf{x}^{\prime}\right)$ term in equation~\ref{eq:kernel}) stretches when atoms approach each other. This makes the covariance function non-stationary with respect to the atom coordinates and allows faster variation of energy in those directions (see more details in Ref.~\cite{Koistinen2019Nudged}).
%
The inverse distance covariance function measures the similarity of two input atomic coordinates $\mathbf{x}$ and $\mathbf{x}^{\prime}$ as \cite{Koistinen2019minimum_mode}: 
\begin{equation}\label{eq:kernel}
\begin{aligned}
&k_{1 / r}\left(\mathbf{x}, \mathbf{x}^{\prime}\right) \\
&\quad=\sigma_{\mathrm{c}}^{2}+\sigma_{\mathrm{m}}^{2} \exp \left(-\frac{1}{2} \underbrace{\sum_{\substack{i \in A_{\mathrm{m}}}} \sum_{\substack{j \in A_{\mathrm{m}}, j>i \\
\mathrm{~V} \\
j \in A_{\mathrm{f}}}} \frac{\left(\frac{1}{r_{i, j}(\mathbf{x})}-\frac{1}{r_{i, j}\left(\mathbf{x}^{\prime}\right)}\right)^{2}}{l_{\phi(i, j)}^{2}}}_{\mathcal{D}_{1/r}\left(\mathbf{x}, \mathbf{x}^{\prime}\right)}\right)
\end{aligned}
\end{equation}
Here, $\mathbf{x}$ (or $\mathbf{x}^{\prime}$) denotes a 3N-dimensional configuration vector including the Cartesian coordinates of the atomic system with N atoms, $\mathbf{x}=[x_{11},x_{12},x_{13},...,x_{N1},x_{N2},x_{N3}]^{\text{T}}$. $r_{i, j}$ is the distance between atoms $i$ and $j$, defined as $r_{i,j}=\sqrt{\sum_{d=1}^{3}\left(x_{id}-x_{jd}\right)^2}$. $l_{\phi}(i, j)$ denotes the length scale for the atom pair $\phi_{(i, j)}$. $\sigma_m$ controls the magnitude of the covariance function, and $\sigma_c$ is the variance of a constant Gaussian prior distribution. The $\mathbf{l}_{\phi}$ vector (with a size of the number of atomic pairs), $\sigma_m$ and, $\sigma_c$ are the training parameters of the inverse distance covariance function.
As shown in equation~\ref{eq:kernel}, index $i$ runs over moving atoms $A_m$ and index $j$ runs over other moving atoms and frozen atoms $A_{f}$. Therefore, atom pairs are only defined between moving atoms and the rest of the moving and frozen atoms. This construct reduces the total number of pairs (i.e., the size of vector $\mathbf{l}_{\phi}$) from $\frac{1}{2}N\times(N-1)$ to $\frac{1}{2}N_m\times(N_m-1)+N_m\times N_f$, where $N$, $N_m$, and $N_f$ denote the number of all atoms, moving atoms, and frozen atoms, respectively. In the examples of this study, we define the diffusing atom to be the moving atom ($N_m=1$), and frozen atoms are those confined in a sphere of radius $r_f$ centered around the moving atom. We call this spherical region the \textit{active region}. We examine the effect of different $r_f$ values on GPR performance in section~\ref{sec:results}. The partitioning of the atomic system into the moving and frozen atoms is specially advantageous in reducing the number of degrees of freedom (i.e., the number of atomic pairs or size of $\mathbf{l}_{\phi}$ vector) for the solid-state phases in our study. 
Additionally, by only including the atomic pair distances between a moving atom and frozen atoms in the covariance function, we inform the GPR model with the most physically important atomic pairs. In other words, the atomic pair distances formed between non-moving atoms carry less physically significant information in describing the potential energy surface. This physical knowledge embedded into the construct of the covariance function aids the model to learn the energy surface more effectively. 
%
%
\vspace{2mm}
\par
\noindent\textbf{\textit{GPR Training \& Prediction}}. For training the GPR, we adopt a multi-task learning approach \cite{Bonilla2008,Swersky2013} as implemented in GPyTorch~\cite{gardner2018gpytorch}, which enables simultaneous learning of energy and forces by sharing information across the prediction tasks. As we show in section~\ref{sec:results}, multi-task learning outperforms derivative observation learning by reducing the GPR training time and error and enhancing its performance and robustness. Through multi-task GPR \cite{Bonilla2008} the inter-task dependencies are learned based solely on the task identities and the observed data for each task, unlike the derivative observation GPR which explicitly enforces the dependence of forces and energy values by equating forces to the negative derivative of energy. For details of the derivative observation GPR learning approach, see equations 28 and 29 of Ref.~\cite{Koistinen2019Nudged}. For multi-task GPR learning, a shared covariance function between tasks $t_1$ and $t_2$ is defined for two inputs $\mathbf{x}$ and $\mathbf{x}^{\prime}$ as \cite{Bonilla2008}:
\begin{equation}\label{eq:kerneljoint}
k\left([\mathbf{x}, t_1],\left[\mathbf{x}^{\prime}, t_2\right]\right)=k_{\mathrm{input }}\left(\mathbf{x}, \mathbf{x}^{\prime}\right) \times k_{\mathrm{task}}(t_1, t_2)
\end{equation}
where $k_{\mathrm{input}}$ is the inverse distance covariance function defined in equation~\ref{eq:kernel} and $k_{\mathrm{task}}$ is the inter-task similarity measure describing the correlation between tasks. In this study, the related tasks are the prediction of energy and atomic force components. The prediction of each force component is a separate task, thus the total number of tasks is $1+3N$ for $N$ atoms in the system. Following the multi-task learning approach of Ref.~\cite{Bonilla2008}, $k_{\mathrm{input}}$ is defined as a “free-form” task-similarity matrix, instead of a parametric covariance function. Specifically, $k_{\mathrm{input}}$ is defined as a positive semi-definite matrix which is approximated by an incomplete-Cholesky decomposition of rank P. Here, we use rank 1 for approximating $k_{\mathrm{input }}$, resulting in only one additional trainable parameter of the GPR. More details about the paramterization of the task-similarity matrix are given in Ref.~\cite{Bonilla2008}. 

Given a set of $M$ tasks and $D$ training data points (or observations), the shared covariance matrix $\mathbf{K} \in \mathbb{R}^{DM \times DM}$ can be expressed as the Kronecker product of the input covariance matrix $\mathbf{K}_{\mathrm{input}} \in \mathbb{R}^{D \times D}$ and the task covariance matrix $\mathbf{K}_{\mathrm{task}} \in \mathbb{R}^{M \times M}$:
\begin{equation}\label{matrixkernel}
\mathbf{K} = \mathbf{K}_\text{task} \otimes \mathbf{K}_\text{input}
\end{equation}

Here, the $D$ distinct observations constitute the training data set $\mathbf\{\mathbf{X},\mathbf{Y}\}$. $\mathbf{X}$ consists of input atomic configurations, $\mathbf{X}=\{\mathbf{x}_1,...,\mathbf{x}_D\}$, and $\mathbf{Y}$ consists of DFT-evaluated energy and force components at $\mathbf{X}$, $\mathbf{Y} = (E_1,..., E_D, \mathbf{f}_1,...,\mathbf{f}_D)^T$, where $E$ and $\mathbf{f}=[f_{1,1},f_{1,2},f_{1,3},...,f_{N,1},f_{N,2},f_{N,3}]$ are the energy and force vector for each input atomic configuration, respectively.
The set of trainable parameters $\theta=\{\sigma_c,\sigma_m,\mathbf{l}_{\phi}\}$ and the single parameter of matrix $\mathbf{K}_\text{task}$ are optimized by maximizing the log marginal likelihood over the shared covariance function \cite{Rasmussen2006}
\begin{equation}
\underset{\theta,K_{task}}{\mathrm{argmax}}\left\{\mathcal{L} = - \frac{1}{2} \mathbf{Y}^T (\mathbf{K} + \sigma^2_n \mathbf{I})^{-1} \mathbf{Y} - \frac{1}{2} \log |\mathbf{K} + \sigma^2_n \mathbf{I}| - \frac{D}{2} \log 2\pi \right\}
\end{equation}
Here, $\mathbf{K}$ is the shared covariance matrix of equation \ref{matrixkernel}, $\mathbf{I}$ is the identity matrix, and $\sigma^2_n$ is the random noise variance, which we set to $10^{-4}$.
%

The GP approximation for the energy or each force component is then obtained as the mean prediction on a new data-point $x^*$ for task $l$ using the posterior distribution conditional on the optimized parameters: 
\begin{equation}\label{gprpred}
f_l(x^*) = (k_{\mathrm{task}}^l \otimes  k_{\mathrm{input}}^*) (\mathbf{K} + \sigma^2_n \mathbf{I})^{-1} \mathbf{Y}
\end{equation}
where $k_{\mathrm{task}}^l$ denotes the $l^\mathrm{th}$ column of $\mathbf{K}_\text{task}$ and $k_{\mathrm{input}}^*$ is the vector of covariances between the query point $x^*$ and the training points. 
\subsection{GPR-accelerated dimer}
In this study, we use the dimer saddle point search method as detailed in Ref.~\cite{ren2013climbing}. A dimer contains a pair of  two auxiliary points in the atomic configuration space of dimension $3N$ (or images), separated by a fixed distance of $0.1$ \AA. Each dimer iteration is divided into a set of rotation steps and a translation step. During the rotation steps, the dimer is rotated around its midpoint to find the orientation that gives the lowest total energy of the two images. This gives the direction of the lowest curvature mode or the minimum mode. The dimer is then translated by reversing the force components in the minimum mode direction multiplied by a step size of $0.1$ \AA. Details of the dimer algorithm used in this work are presented in Ref.~\cite{ren2013climbing}. We used Algorithm B1 and Algorithm B3 of Ref.~\cite{ren2013climbing}, respectively, for rotation and translation of the dimer. 

For a GPR-guided dimer, the GPR-approximation of the energy and force components (according to equation~\ref{gprpred}) are used to provide the forces acting on the images of the dimer during rotation and translation. Rotational forces are then defined according to the projected atomic forces on the two images of the dimer as $F_R$. Rotational steps are carried out until $F_R$ falls below a threshold (0.1 eV/\AA) or a maximum number of rotations are performed. The maximum number of rotations is set to 5 for a standard DFT-dimer and to 25 for a GPR-guided dimer. We use the conjugate gradient algorithm for determining the rotational plane of the dimer. As explained in the previous section, the GPR model is updated (retrained) after every $n_h$ translation steps by using an expanded set of DFT-calculated training data points. The final convergence of the dimer to the saddle point is achieved when the maximum atomic force approximated by the GPR is below $0.01$ eV/atom. An accurate DFT-calculated force at the final point of dimer is used to confirm the convergence. 
\section{Results}\label{sec:results}
\subsection{Validation: 2-Dimensional Sinusoidal Potential Model}
\begin{figure}[!h]
\centering
\includegraphics[width=0.7\textwidth]{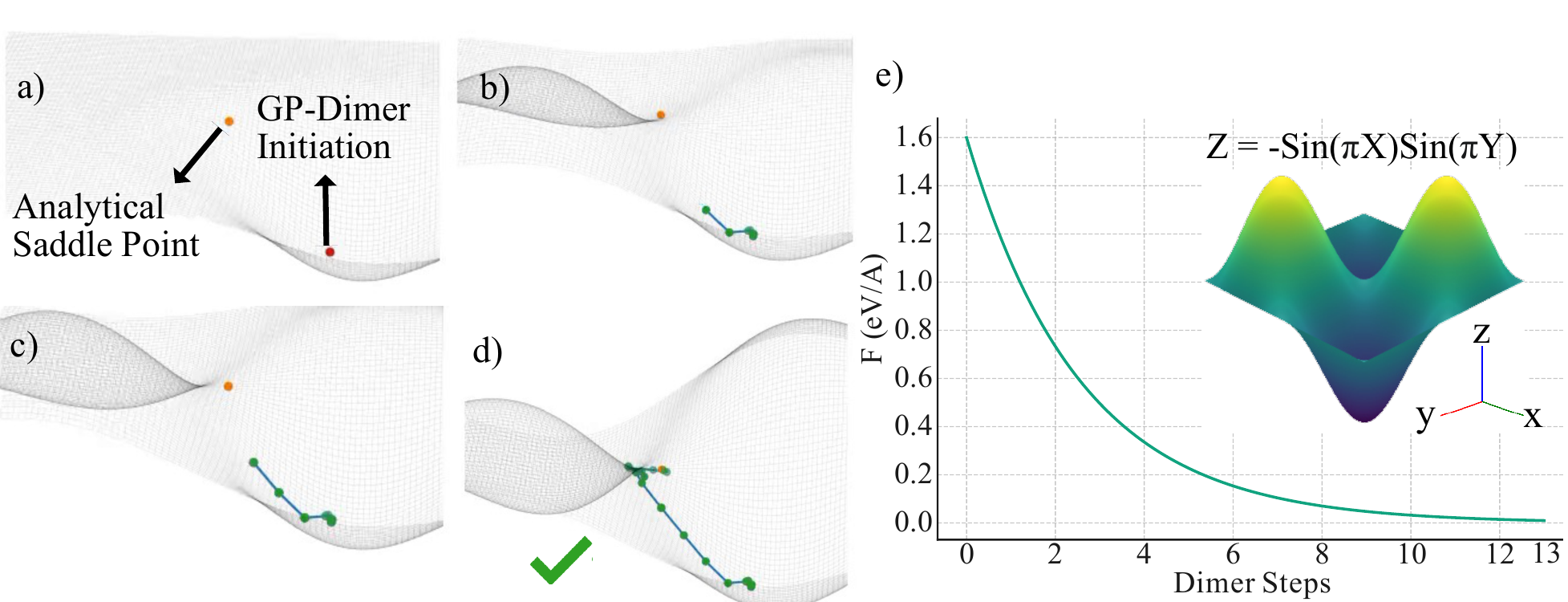}
\caption{The evolution of the GPR approximation of the potential energy surface and the dimer location on the GPR-approximated energy surface at (a) the initial point, and after (b) 4, (c) 5 and (d) 13 dimer translation steps. (d) The final step of the GPR-dimer, where it reaches the exact saddle point at x=0 and y=0. (e) Force magnitude (i.e., derived from the negative gradient of the GPR-approximated potential energy) versus the GPR-dimer steps. The inset illustrates the functional form of the 2D potential model. } 
\label{fig:toy}
\end{figure}

We first validate the GPR-dimer method of this work, as detailed in section~\ref{sec:method}, on a toy potential model of sinusoidal form. The model has the function form of $z=-\sin{\pi x}\sin{\pi y}$, where $z$ denotes the potential energy value and $x$ and $y$ constitute the two coordinates of the input, mimicking the atomic coordinates in a 2-dimensional space. We initiate the GPR-dimer from the minimum on the potential surface at $x=-0.5$ and $y=-0.5$. The GPR-approximated energy surface $\Tilde{z}$ is updated (or retrained) after each dimer translation step (i.e., $n_h=0$). The training set is expanded by a new data point, $z_i(x_i,y_i)$, at each dimer translation step, where $(x_i,y_i)$ specify the 2D atomic coordinates of the new dimer location, and then the GPR is trained on the new training data. The threshold for training of the GPR is for the mean absolute error (MAE) of the force and energy to drop below 0.01 eV/\AA ${}$ and 0.01 eV, respectively. Figure~\ref{fig:toy} illustrates the evolution of the GRP energy surface, $\Tilde{z}$ , and the dimer walker location at different dimer translation steps. Fig.~\ref{fig:toy} also shows the decrease of the force magnitude, $|\Tilde{F}|$ (i.e., $|\Tilde{F}|=\sqrt{\Tilde{F}_x^2+\Tilde{F}_y^2}$ where $\Tilde{F}_x=-\frac{\partial \Tilde{z}}{\partial x}$ and $\Tilde{F}_y=-\frac{\partial \Tilde{z}}{\partial y}$) as the GPR-dimer progresses toward the saddle point. The GPR-dimer reaches the saddle point at $x=0$ and $y=0$ after 13 translation steps, with a total number of 60 dimer rotations. For the 2D potential model, we use the radial basis covariance function (RBF), as implemented in GPyTorch~\cite{gardner2018gpytorch} and the derivative-observation learning approach~\cite{SolakAdvances2002}.
\subsection{Validation: Self-Diffusion in \textit{bcc} Mo}
\begin{figure}[!h]
\centering
\includegraphics[width=0.7\textwidth]{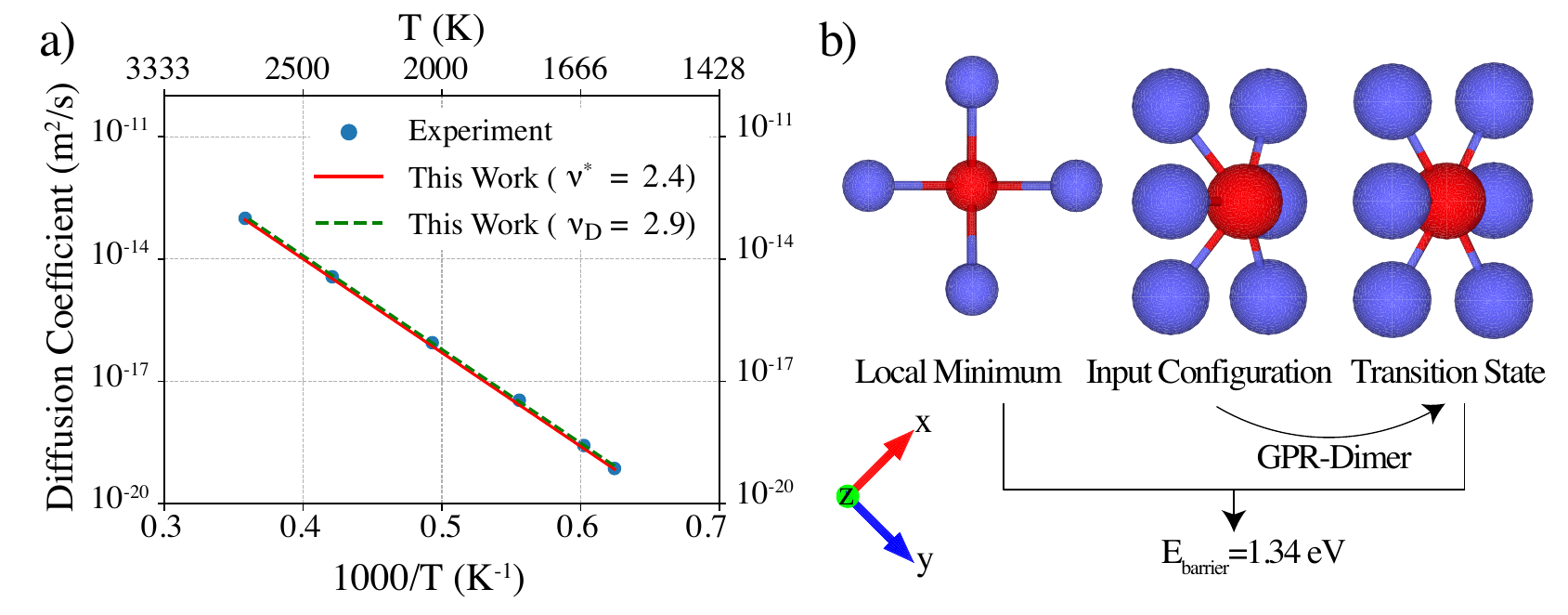}
\caption{a) Diffusion coefficient versus temperature in \textit{bcc} Mo. The predictions of this work, from DFT calculations and the GPR-dimer saddle point search method, are compared against experimental measurements~\cite{herzig1987}. b) The atomic configuration in the vicinity of the diffusing atom (colored by red) in \textit{bcc} Mo for (left) the local minimum state (vacancy residing on a \textit{bcc} site), (middle) the input atomic configuration to the GPR-dimer, and (right) the saddle point on the energy surface or the transition state, output from the GPR-dimer algorithm.}
\label{fig:DiffusionMo}
\end{figure}

To validate that the GPR-dimer method can successfully identify the transition state of a solid-state process, we apply it to calculate the energy barrier for a vacancy diffusive hop in the \textit{bcc} phase of Mo. We initiate the GPR-dimer walker at an atomic configuration that is a linear interpolation between the initial state (a local minimum state), where a \textit{bcc} lattice site is vacant, and the final state (a symmetrically-equivalent local minimum state), where the vacant \textit{bcc} site has hopped to the nearest neighbor. The input configuration is a weighted average of the atomic coordinates with 3/4 contribution from the initial state and 1/4 from the final state (see Figure~\ref{fig:DiffusionMo}(b)). Supplementary Note 1 examines the GPR-dimer application using the local minimum configuration as the initial input. Given the input configuration, we perform a standard DFT-dimer for two translation steps to provide the data points for training the GPR. A total of three atomic configurations are used to train the GPR ($n_i=3$). The GPR-dimer is then launched to locate the saddle point. We use the inverse distance covariance function with an active region of size $r_f=3$ \AA${}$. For training the GRP, we employ the multi-task learning as detailed in section~\ref{sec:method}. The training data set is expanded by an additional atomic configuration at every 10$^\text{th}$ translation step of the dimer (i.e., $n_h=10$), followed by an update (or retraining) of the GPR. The criterion of reaching the saddle point is for the total force magnitude to be less than 0.01 eV/\AA${}$, where the GPR-dimer stops. The total force magnitude is calculated as $F=\sqrt{\sum_{i=1}^{N}{F_{i1}^2+F_{i2}^2 + +F_{i3}^2}}$. The energy difference between the final step of the GPR-dimer (or the transition state) and the initial state of the vacancy hop (the local minimum) is calculated to provide the energy barrier for the vacancy diffusive hop (or the enthalpy of vacancy migration), $\Delta H_m$. Energies of the transition and local minimum states are both calculated using DFT. The calculated enthalpy of vacancy migration is equal to 1.34 eV (see Figure~\ref{fig:DiffusionMo}(b)), which is in good agreement of our previous calculation using NEB \cite{Kadkhodaei2022}.  

Using the calculated enthalpy of vacancy migration, $\Delta H_m$, we validate the diffusion coefficient of \textit{bcc} Mo as a function of temperature with experimental measurements~\cite{herzig1987}. Figure \ref{fig:DiffusionMo}(a) shows the calculated diffusion coefficient based on the located saddle point on the energy surface by the GPR-dimer method in comparison with experimental results~\cite{herzig1987}. We calculate the self-diffusion coefficient for monovacncy diffusive jumps according to $D=C_v d^2 \Gamma$, where $d$ is the vacancy (or atom) jump distance, $C_v$ is the equilibrium vacancy concentration, and $\Gamma$ is the successful vacancy jump rate. Vacancy jump distance in \textit{bcc} is equal to the nearest neighbor distance or $\frac{\sqrt 3}{2}a_0$, where $a_0$ is the lattice constant. Vacancy concentration at temperature $T$ is given by $\displaystyle C_v = \exp(\frac{\Delta S_f}{k_B})\exp(-\frac{\Delta H_f}{k_BT})$, where $\Delta H_f$ and $\Delta S_f$ are the formation enthalpy and entropy of vacancy, respectively, and $k_B$ is the Boltzmann constant. We obtain the DFT-calculated values of $a_0$ and $C_v$ from our previous results in Ref.~\cite{Kadkhodaei2022}. The vacancy jump rate $\Gamma$ is obtained from the migration enthalpy, $\Delta H_m$, and the effective vibration frequency along the migration path, $\nu^*$, by $\Gamma = \nu^* \exp(\frac{-\Delta H_m}{k_BT})$. Here, $\Delta H_m$ is the vacancy migration energy barrier calculated according to the saddle point located by the GPR-dimer method. We obtain the DFT-calculated $\nu^*$ from Ref.~\cite{Kadkhodaei2022}, which calculates $\nu^*$ as the ratio of the product of normal vibration frequencies of the initial state of atomic migration, $\nu_i$, to that of the non-imaginary normal frequencies of the transition state, $\nu^{\prime}_j$, i.e., $\nu^* = \frac{\prod^{3N-3}_{i=1} \nu_i}{ \prod^{3N-4}_{j=1} \nu^{'}_j}$. Alternatively, we estimate $\nu^*$ to be equal to the Debye frequency, $\nu^*\approx\nu_\text{D}$, which is calculated from Debye temperature $\Theta_\text{D}$ as $\nu_\text{D}=\Theta_\text{D} \frac{k_\text{B}}{\hbar}$, where $\hbar$ is the reduced Plank's constant. 
\subsection{Validation: Sulfur Diffusion in \textit{hex} \ce{MoS2}}
To further validate the accuracy of the GPR-dimer method in identifying transition states in solid-state processes, we apply it to calculate the energy barrier for the monovacancy-sulfur diffusive jumps in \textit{hexagonal} \ce{MoS2} (P\ce{6_3}/mmc space group with 2b and 4f wyckoff positions for Mo and S respectively). 

\begin{figure}[!h]
\centering
\includegraphics[width=0.6\textwidth]{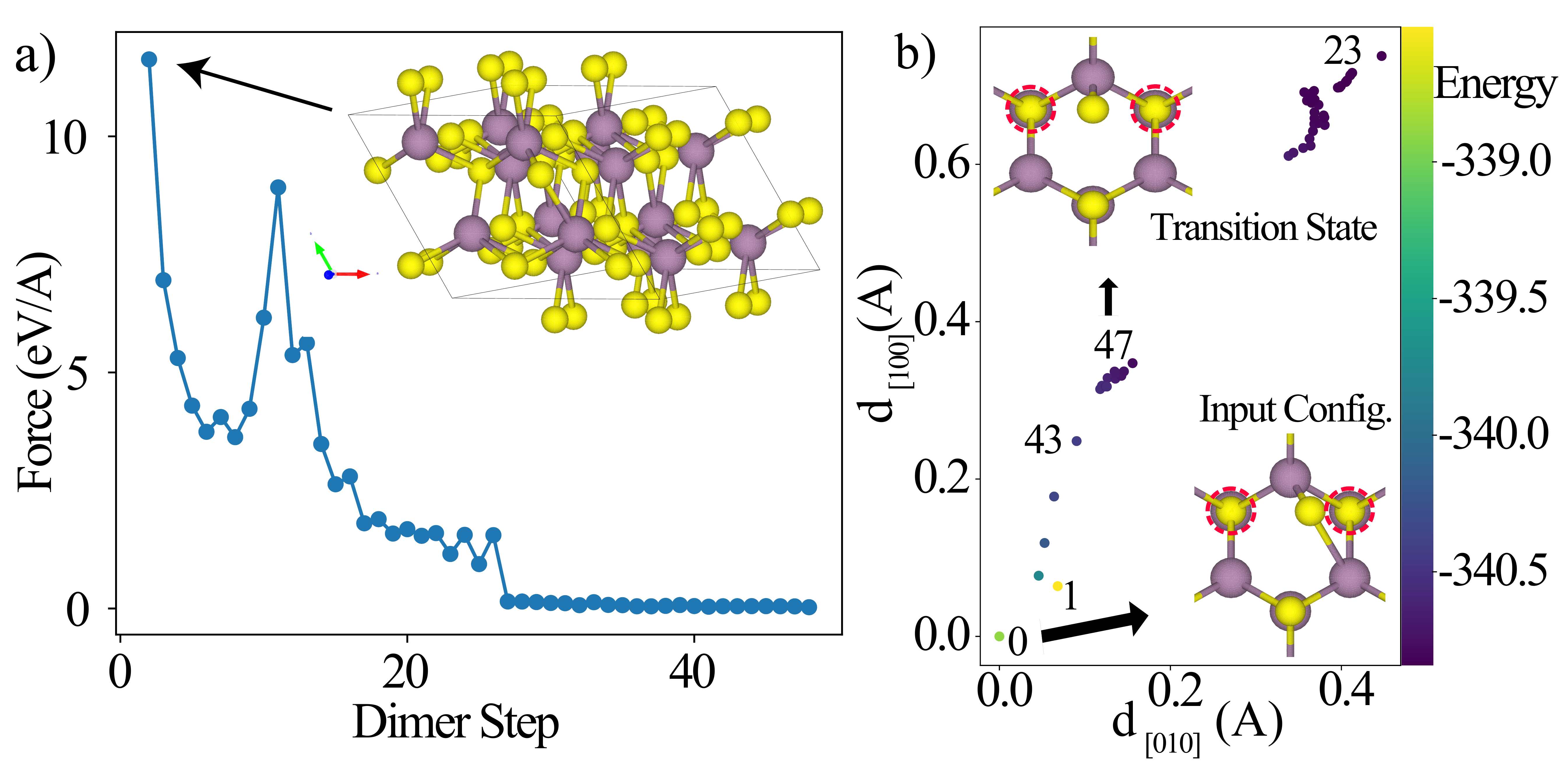}
\caption{Evolution of the GPR-dimer to locate the transition state of sulfur-monovancy diffusive jump in \textit{hex} \ce{MoS2}. a) The total force magnitude of the atomic configuration versus dimer translation step. The inset depict the input configuration to the GPR-dimer. b) Displacement of the diffusing sulfur atom during the GPR-dimer saddle point search, projected along the [100] and [010] directions. The insets depict the atomic configuration in the vicinity of diffusing sulfur. Sulfur, molebdynum, and vacancy are colored by yellow, purple, and dotted red circles, respectively. The color bar maps the energy of different dimer steps.}
\label{fig:DiffusionMoS2}
\end{figure}
The input atomic coordinates to the GPR-dimer are linearly interpolated with 3/4 contribution from the local minimum configuration, where one sulfur site is vacant, and 1/4 from the final state, where the vacancy and the nearest sulfur has exchanged their positions (see Figure~\ref{fig:DiffusionMoS2} for the input configuration).  
Subsequently, we launch the GPR-dimer method to locate the saddle point along the diffusion pathway. Like the example of \textit{bcc} Mo, we use the first three atomic configurations from the translation steps of the DFT-dimer to train the GPR (i.e., $n_i=3$), and we consequently update (or retrain) the GPR after every 10 translation steps (i.e., $n_h=10$). The use an active region of radius $r_f=5 \AA$ centered at the diffusing sulfur atom for the inverse-distance covariance function of equation~\ref{eq:kernel}. Figure~\ref{fig:DiffusionMoS2}(a) shows the evolution of the total force magnitude of the atomic configuration at different translation steps of the GPR-dimer. Once the total force is below 0.01 eV/\AA, we stop the GPR dimer. Figure~\ref{fig:DiffusionMoS2}(b) shows the displacement of the diffusing atom projected on the [100] and [010] directions during the GPR-dimer evolution. The energy of the output of the GPR-dimer (or the transition state) and the local minimum configuration are calculated using DFT. The calculated energy difference between these two configurations equals 2.3 eV, which is in good agreement with the DFT-NEB calculation of Ref.~\cite{DFT_MoS2}.
\subsection{GPR-Dimer Performance Analysis}
In this section, we examine the impact of various parameters on the performance of the GPR-dimer method introduced in this study. These parameters include: 1) learning approach or GPR training method, 2) energy surface sampling frequency or number of dimer translation hops between DFT-calculations, 3) size of the active region containing frozen atoms in the inverse-distance covariance function (Equation~\ref{eq:kernel}), and 4) number of preceding DFT-sampled data points included in GPR training set at each retraining step. We showcase our investigations for the case of a monovacancy diffusive jump in \textit{bcc} Mo. 

\begin{figure}[!h]
\centering
\includegraphics[width=1.0\textwidth]{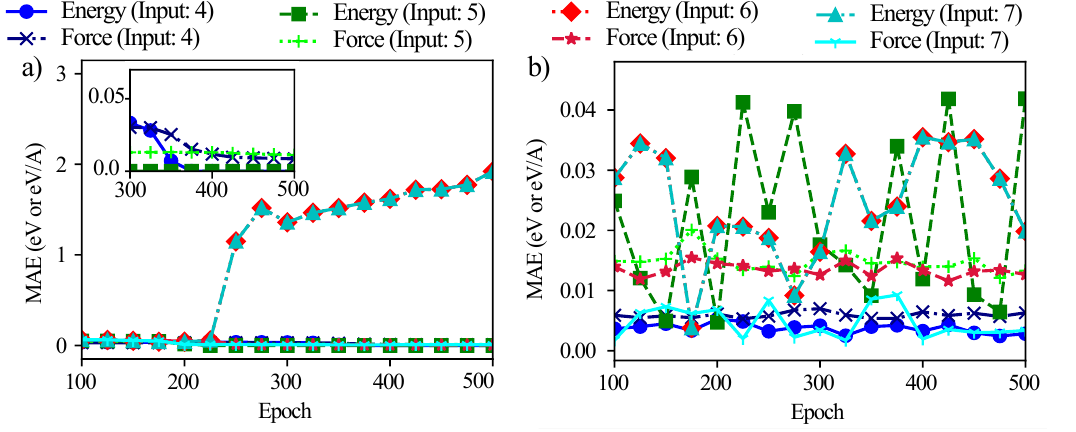}
\caption{Comparison of gradient observation and multi-task learning on the GPR predictive performance. The mean absolute error (MAE) of the energy prediction (eV) and the atomic force magnitude prediction (eV/\AA) in terms of the training epoch for a) gradient observation and b) multi-task learning. Different line styles and markers indicate various sizes of the training data.}
\label{fig:GO_vs_multi}
\end{figure}

To evaluate the impact of different GPR learning approaches, we compare multi-task learning (explained in Section~\ref{sec:method}) with the commonly used derivative-observation learning. Figure~\ref{fig:GO_vs_multi} illustrates the learning curves for multi-task learning and derivative-observation learning. The GPR serves as the initial surrogate model for the energy surface, and the training data consist of $n_i$ consecutive DFT-calculated dimer translation steps starting from the initial atomic configuration (as detailed in Section~\ref{sec:method}).
The mean absolute error (MAE) for energy and force predictions are presented over the training epoch for various numbers of input training data, $n_i$. The multi-task learning method demonstrates superior learning performance in terms of prediction accuracy and stability compared to derivative-observation learning. The multi-task learned GPR consistently exhibits lower MAE across all epochs and input data sizes. Furthermore, the MAE remains low and relatively stable for different input data sizes, indicating the robustness of multi-task learning and its lower sensitivity to observations (or training data).
In contrast, the GPR trained with derivative-observation learning shows an increase in MAE as the number of input data increases. Specifically, the MAE for energy prediction starts to rise significantly above 1 eV after around 200 epochs, while the MAE for force prediction remains low. This behavior is likely attributed to over-fitting of the derivative-observation GPR to a single task. The explicit relationship between forces and energies in derivative-observation learning constrains the optimization process, making it prone to issues such as over-fitting to forces in this particular example.
On the other hand, multi-task learning employs an implicit regularization effect, mitigating the risk of over-fitting to a single task. By simultaneously learning multiple correlated tasks, the model captures common latent features that generalize better to new data.

\begin{figure}[htbp]
\centering
\includegraphics[width=\textwidth]{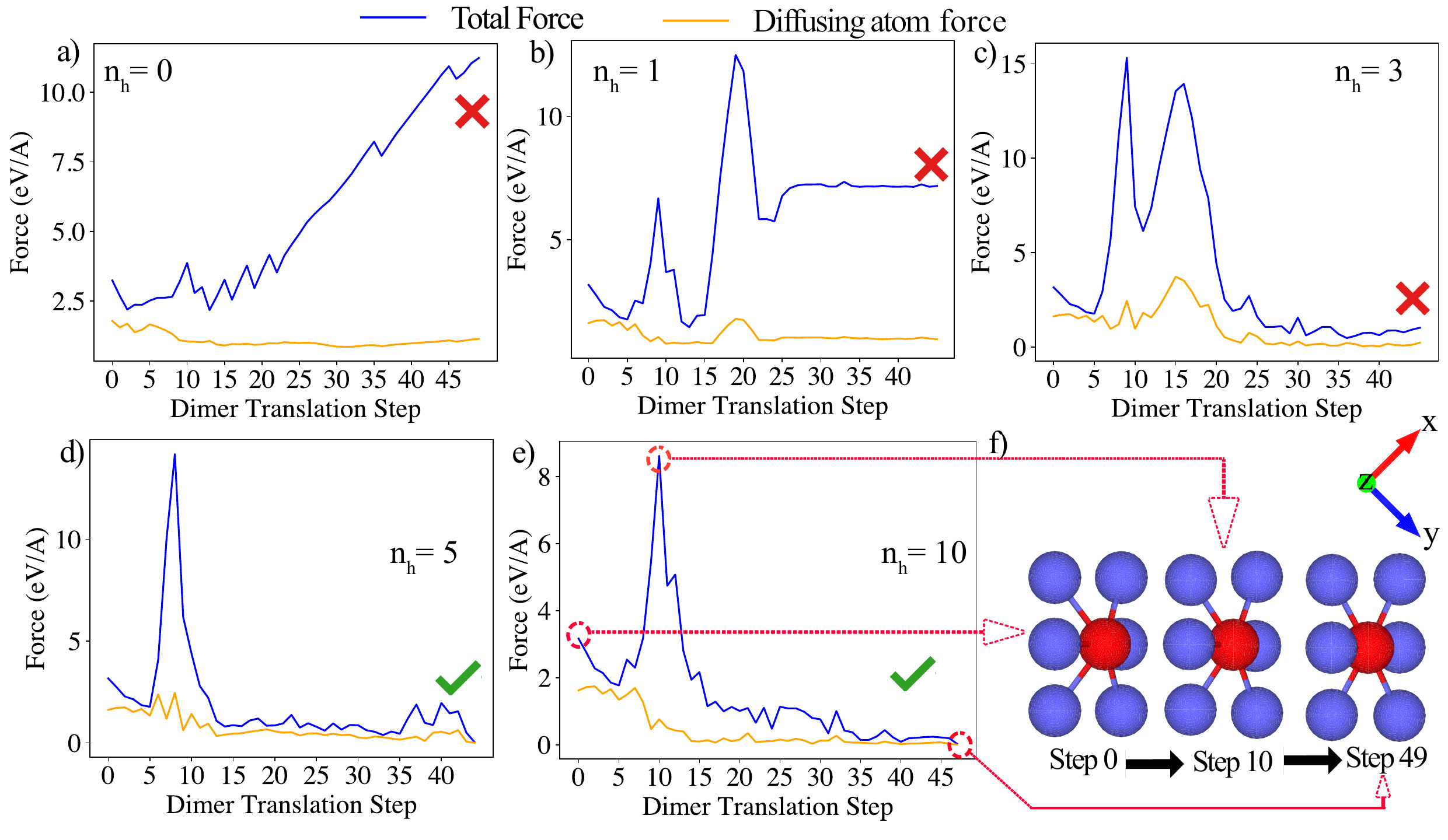}
\caption{Assessing the performance of the GPR-dimer for different translation hop values in the translation-hop sampling approach. The evolution of the total atomic force and the diffusing atom force in terms of dimer translation steps for a) zero, b) one, c) three, d) five, and e) ten translation hops. The x-axis represents the GPR-dimer translation steps only where the GPR is updated with new DFT-calculated data points. f) Evolution of the atomic configuration in the vicinity of the diffusing atom at different steps for the GPR-dimer with $n_h=10$.}
\label{fig:hop}
\end{figure}

We assess the impact of the number of translation hops, denoted as $n_h$, in the translation-hop sampling approach described in Section~\ref{sec:method}, on the performance of GPR-dimer. In Figure~\ref{fig:hop}, we present the behavior of the GPR-dimer for different values of $n_h$ ($n_h={0,1,3,5,10}$). The figure illustrates the DFT-calculated total atomic force magnitude and the force magnitude of the diffusing atom as a function of the dimer translation step. 
The force values are only displayed for the dimer translation steps where the GPR is updated, corresponding to the atomic configurations associated with the DFT-calculated steps. The GPR-predicted forces at the intermediate steps between DFT sampling points are not shown.
Based on our examination, we observe that the GPR-dimer with zero, one, and three translation hops failed to locate the saddle point. In contrast, the GPR-dimer with five and ten translation hops successfully converged to the saddle point. This observation provides valuable insights into how the sampling frequency along the dimer path influences the GPR surrogate model, striking a balance between a localized and global representation of the energy surface.
For zero, one, and three translation hops, the GPR becomes excessively influenced by the energy surface in the vicinity of the dimer path. Consequently, it overfits to the minimum-mode following path while neglecting the broader energy landscape. On the other hand, delaying the sampling by five or ten translation hops enables the GPR to capture a more balanced representation, encompassing both the vicinity of the path and the wider energy surface shape. In other words, the translation-hop sampling approach provides the opportunity to balance exploration and exploitation through tuning the $n_h$ parameter.  
It is worth noting that in the case of a 2D potential model (depicted in Figure~\ref{fig:toy}), no translation hops are required. The GPR demonstrates robustness against overfitting due to the low dimensionality of the input space of the covariance function.

\begin{figure}[!h]
\centering
\includegraphics[width=1.0\textwidth]{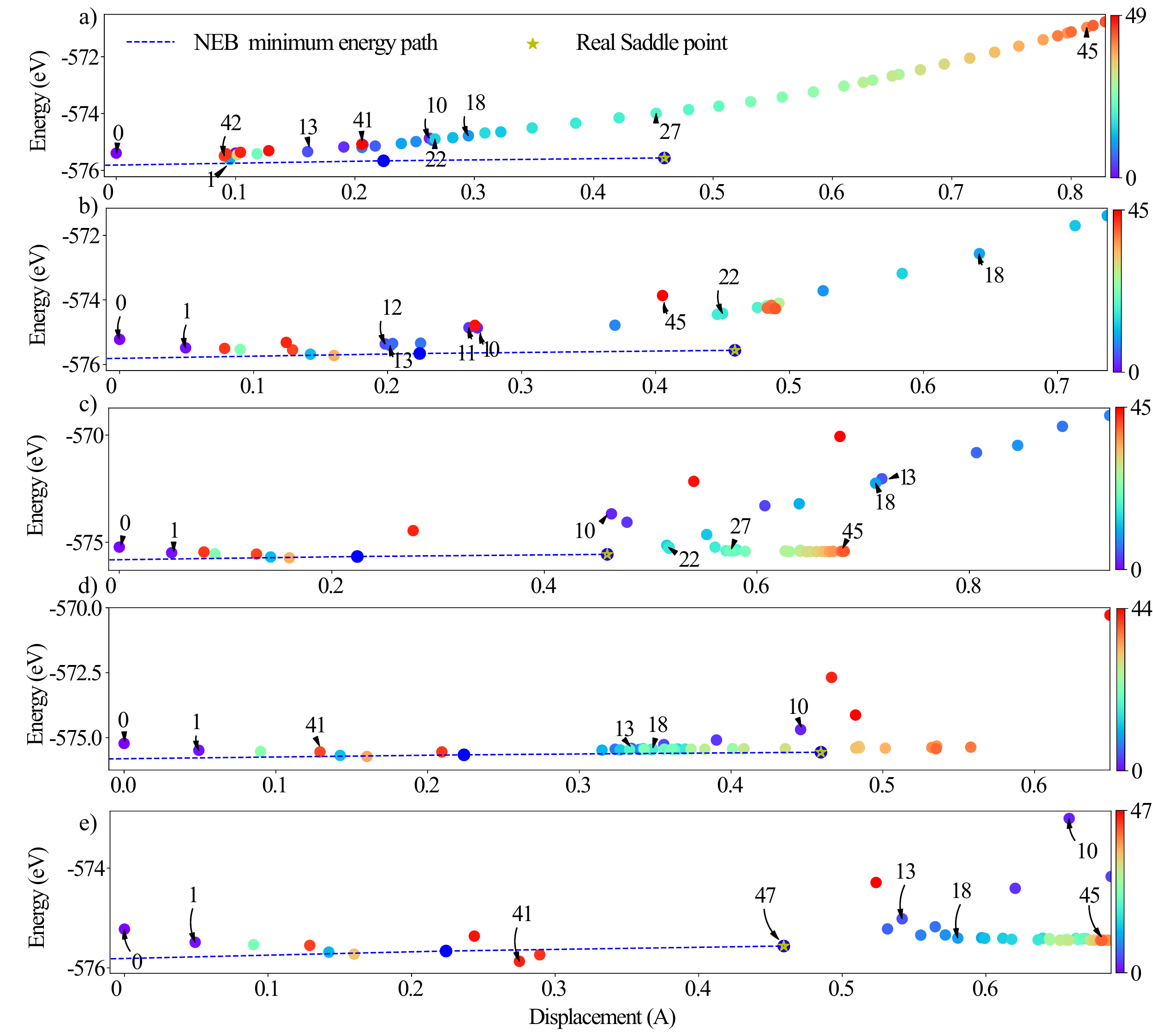}
\caption{The energy profile of the GPR-dimer for different translation hops. Energy of the GPR-dimer versus its displacement along [111] direction for a) zero, b) one, c) three, d) five, and e) ten translation hops. The color bar depict the progression of the GPR-dimer walker.}
\label{fig:hop_walker}
\end{figure}

To further investigate the influence of translation-hop sampling frequency on the GPR model, we present the energy profile of the GPR-dimer for different translation hops in Figure~\ref{fig:hop_walker}. The minimum energy pathway, obtained from NEB calculations implemented in VTST~\cite{Henkelman2000climbing}, is shown as a reference. The movement of the GPR-dimer is projected along the minimum energy path direction (or along the [111] lattice direction), which serves as the x-axis in Figure~\ref{fig:hop_walker}.
In the case of zero hops (Figure~\ref{fig:hop_walker}(a)), the GPR-dimer bypasses the saddle point and explores high-ridge regions of the energy surface. This is because the GPR is over-fitted to the walker pathway and most likely to the noise in the initial dimer walker oscillations, which results in misguiding the walker. For one or three hops (Figure~\ref{fig:hop_walker}(b) and (c)), the walker goes back and forth between lower and higher energy regions but fails to locate the saddle point.
In contrast, for five and ten hops (Figure~\ref{fig:hop_walker}(d) and (e)), the GPR-dimer walker deviates from the NEB path at the beginning, as the GPR's accuracy in predicting energies near the pathway is reduced. The walker takes larger steps and explores a diverse range of points on the energy surface, eventually converging to the saddle point as it progresses and incorporates more sampled data points. Supplementary Figure S1 illustrates a 2D projection of the GPR-dimer trajectory on the high-dimensional energy surface for different translation hops.

\begin{figure}[!h]
\centering
\includegraphics[width=1.0\textwidth]{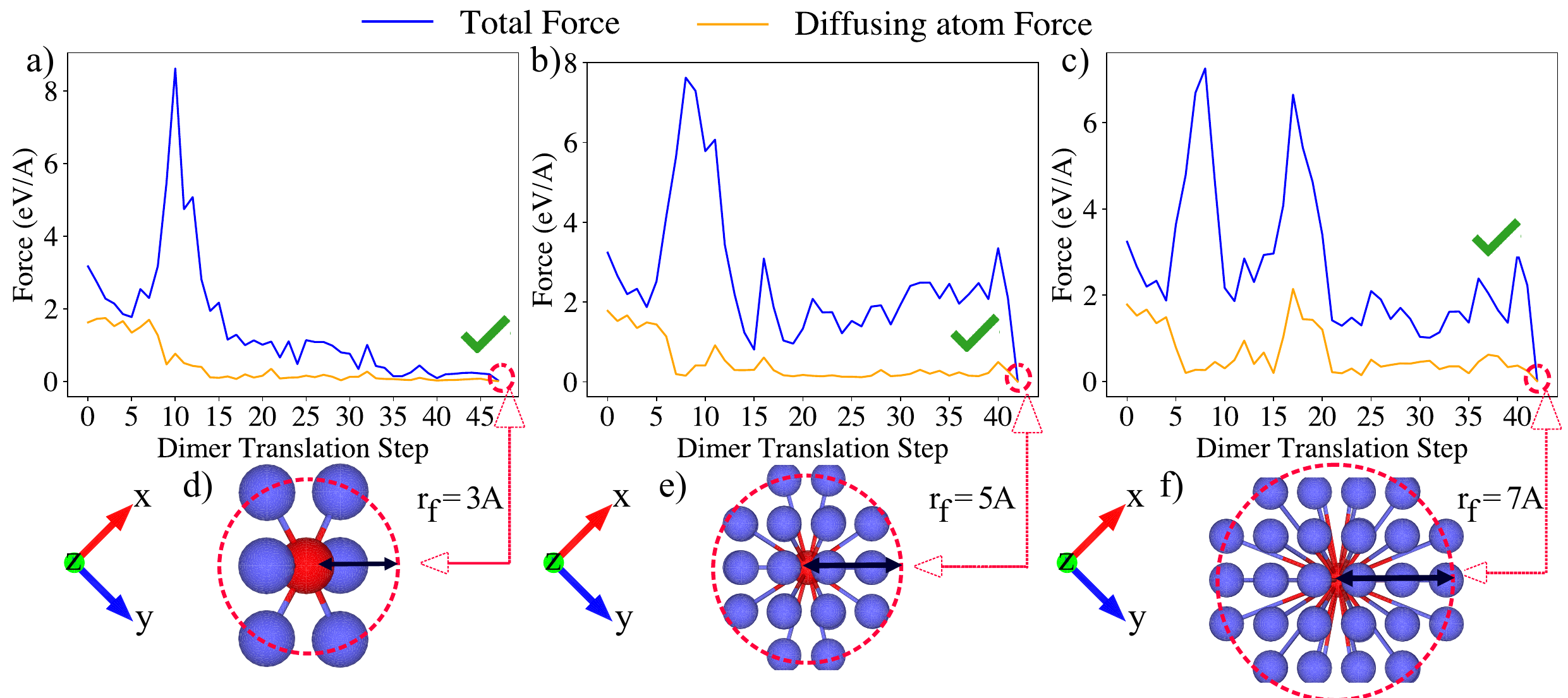}
\caption{The impact of active region radius on GPR-dimer performance. DFT-calculated total atomic force and diffusing atom force for radii of a) 3 \AA, b) 5 \AA, and c) 7 \AA. The x-axis represents the GPR-dimer translation steps at which GPR is updated (i.e., DFT calculation is performed). d-f) The atomic configuration surrounding the diffusing atom for the final GPR-dimer step in each case. }
\label{fig:active}
\end{figure}

We investigate the impact of the active region radius, denoted as $r_f$ (Section \ref{sec:method}), on the overall performance of the GPR-dimer. Increasing the active region radius results in more frozen atoms in the inverse distance covariance function (equation ~\ref{eq:kernel}). 
This leads to more pairs between the moving atom (or the diffusing atom) and frozen atoms, providing the GPR with more information about the surrounding atomic configuration.
Figure~\ref{fig:active} illustrates the GPR-dimer's behavior for $r_f$ values of 3, 5, and 7 \AA${}$ (associated with the first, second, and third nearest neighbors of the moving atom, respectively), with a fixed translation hop of $n_h=10$. The DFT-calculated total atomic force and diffusing atom force are shown as a function of the GRP-dimer step.
In all three cases, the GPR-dimer successfully locates the saddle point. However, as shown in Figure~\ref{fig:active}, the force evolution is smoother for $r_f=3$ \AA${}$ compared to larger active regions. The force exhibits an early peak and a monotonic decrease as the GPR-dimer progresses towards the saddle point. In contrast, for $r_f=5\AA$ and $r_f=7\AA$, the force demonstrates significant oscillations throughout the GPR-dimer progression, with $r_f=7\AA$ displaying two force peaks before reaching the saddle point.
The smooth evolution of the GPR-dimer observed at $r_f=3 \AA$ provides valuable insights into achieving an optimal balance between the number of atomic pairs incorporated in the GPR's covariance function and the captured physical information. By setting $r_f=3 \AA$, the covariance function effectively captures variations in the pair distance between the diffusing atom and its first nearest neighbors throughout the progression of the GPR-dimer.
Constraining the pair distance information to the first nearest neighbors proves to be the most effective approach for constructing a surrogate model. This is because the pair distances among the nearest neighbors contain the most relevant physical information while maintaining a relatively low number of pairs. Consequently, this results in a smaller size of the $\mathbf{l}_{\phi}$ vector, reducing the risk of overfitting and improving the model's performance.

\begin{figure}[!h]
\centering
\includegraphics[width=1.0\textwidth]{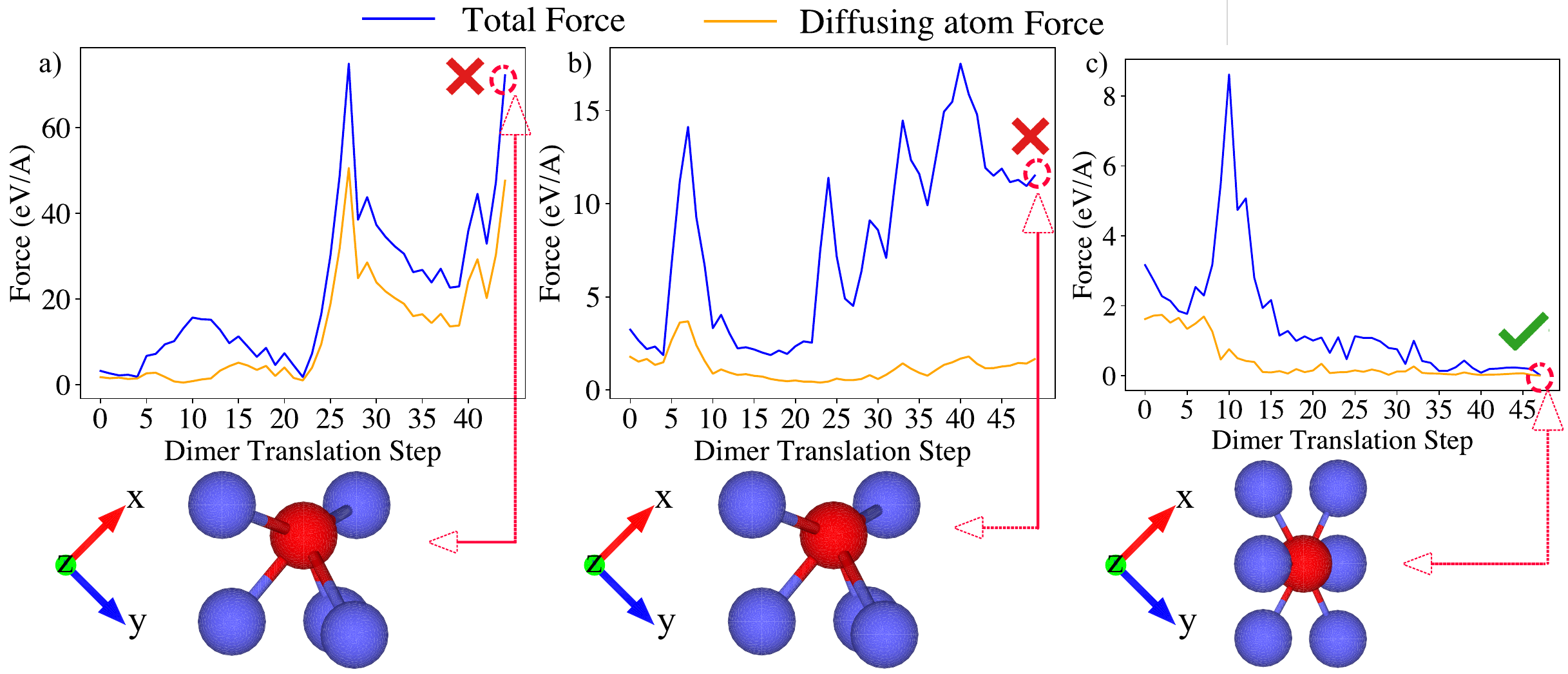}
\caption{Effect of GPR-dimer tail size (or the number of preceding dimer translation steps incorporated in the training set) on the GPR-dimer performance. The DFT-calculated total force and diffusing atom force are shown for a tail size of a) 5, b) 10, and c) all preceding dimer steps. The x-axis shows the GPR-dimer translation steps at which the GPR is updated. The atomic configuration around the diffusing atom is shown for the last GPR-dimer step for each case.}
\label{fig:tail}
\end{figure}

Lastly, we investigate the influence of the GPR-dimer's training data history on its performance. The GPR-dimer's tail size is defined as the last $n_t$ translation steps preceding the current step, which are used as training data for updating the GPR surrogate model. We consider three tail sizes: $n_t=5$, $n_t=10$, and $n_t$ including all preceding DFT-sampled steps. The translation hop is set to $n_h=10$, indicating 10 dimer translations between consecutive DFT-sampled steps. The active region radius is fixed at 3 Å ($r_f=3 Å$).
Figure~\ref{fig:tail} presents the GPR-dimer evolution for different tail sizes $n_t$. The DFT-calculated total atomic force magnitude and diffusing atom force magnitude are shown as a function of the GPR-dimer progression. The GPR-dimer successfully locates the saddle point only when all preceding DFT-sampled steps are included in the training set. This observation aligns with our previous analysis of the translation-hop frequency. Limiting the GPR training data to the last 5 or 10 preceding steps results in a surrogate model representing a local view of the energy surface, causing the walker to bypass the saddle point and move towards high-energy ridges. This is evident from the large force magnitudes observed in Figure~\ref{fig:tail} (a) and (b). Conversely, updating the GPR using the entire history of the dimer walker allows the model to capture a broader view of the energy surface. Our examination reveals that excluding atomic configurations with large repulsive forces (force peaks in the early steps of the GPR-dimer) prevents the GPR from gaining a comprehensive understanding of the energy surface. Therefore, updating the GPR with a diverse set of low- and high-energy points sampled along the walk is necessary for the successful identification of the saddle point by the GPR-dimer.

\subsection{Assessing Computational Efficiency}
In order to evaluate the potential computational efficiency gains achievable by employing a GPR surrogate model with the dimer method, we compare the computational efforts required to locate the saddle point for self-diffusion in \textit{bcc} Mo using three different approaches: the standard DFT-dimer as implemented in our work, the GPR-dimer as implemented in our work, and the DFT-dimer implementation of the Transition State Tools for VASP (VTST)~\cite{Henkelman2000climbing}. Table \ref{tab:comparison} presents a comparison of the total number of DFT calculations of energy and forces necessary for each method to converge to the saddle point. In both the standard dimer methods (our implementation and the VTST code), DFT calculations are performed during both the translation and rotation steps. However, for the GPR-dimer, forces are obtained from the GPR surrogate model during rotations, and DFT force calculations are only carried out after five dimer translation steps when the GPR is updated ($n_h=5$). The GPR surrogate model is initially trained on three DFT-calculated configurations (i.e., $n_i=3$). 
Table \ref{tab:comparison} also includes the number of core-hours required for the DFT calculations by each method, conducted on 2 AMD EPYC 7742 CPUs with 64 cores. As illustrated in Table \ref{tab:comparison}, the GPR-dimer approach entails 44 DFT calculations (equivalent to 17.89 core hours), compared to 54 (20.81 core hours) in our implementation of the standard DFT-dimer. Consequently, utilizing the GPR as a surrogate model yields approximately a 15\% enhancement in computational effort. The VTST implementation, on the other hand, necessitates 60 DFT calculations (equivalent to 23.1 core hours). The slightly higher computational effort associated with the VTST implementation, compared to our implementation of the standard dimer, is likely due to the more efficient conjugate gradient algorithm employed in our work. Specifically, while the VTST code utilizes the original conjugate gradient (CG) algorithm \cite{Henkelman1999dimer}, our code adopts another version of CG \cite{ren2013climbing}. Supplementary tables S1 and S2 presents the computational effort for the GPR-dimer method for different translation hops, $n_h$, and active region radii, $r_f$, respectively. 
%
%
%
\begin{table}[!h]
    \centering
    \begin{tabular}{lll}
    \toprule
        ~ & \textbf{DFT Calculations}  & \textbf{Computational Time (Core-Hour)} \\
    \midrule
        Dimer (this work)  & 54  & 20.81 \\
        GP-Dimer (this work)  & 44 & 17.89 \\
        VTST Dimer  & 60  & 23.1 \\
    \bottomrule
    \end{tabular}
    \caption{Comparison of computational effort between GP-dimer and standard DFT-dimer by the number of DFT calculations and computation time.}
    \label{tab:comparison}
\end{table}

\section{Conclusion}\label{sec:discussion}
We present a methodology that leverages Gaussian process regression (GPR) to develop a surrogate model for the \textit{ab initio} energy surface. By integrating the dimer method with GPR in an iterative feedback loop, we simultaneously sample the energy surface and converge to the saddle
point. The versatility of our proposed GPR-dimer method is demonstrated through its successful application in identifying transition states of vacancy-mediated diffusion in both \textit{bcc} molybdenum and hexagonal molybdenum disulfide.
Our results indicate the promising potential of the GPR-dimer method in enhancing the efficiency of saddle point search in solid-state materials characterized by a large number of atoms. 

To establish a robust and computationally efficient GPR-dimer scheme, we introduced two key components: 1) multi-task GPR learning and 2) translation-hop sampling of training data. The translation-hop sampling approach proves to be essential in striking a delicate balance between exploration and exploitation of the \textit{ab initio} energy surface during the search for the saddle point. This approach enables effective utilization of the available training data while efficiently exploring the energy landscape. Furthermore, by applying the GPR-dimer method to solid-state materials with a high degree of atomic freedom, our findings offer valuable strategies to tackle the challenge of high-dimensionality when employing GPR. 

In summary, our methodology showcases the potential of GPR-dimer as a powerful tool for enhancing saddle point search in solid-state materials. By integrating GPR with the dimer method and incorporating novel strategies, we pave the way for more efficient exploration of complex energy landscapes in the search for transition states.

\section*{Acknowledgement}
This work was supported by the US National Science Foundation Award No. DMR-1954621. This work used Bridges2 at  Pittsburgh Supercomputing Center (PSC) through allocation MAT200013 from the Advanced Cyberinfrastructure Coordination Ecosystem: Services \& Support (ACCESS) program, which is supported by National Science Foundation grants \#2138259, \#2138286, \#2138307, \#2137603, and \#2138296.


\end{document}


\maketitle

\begin{figure}[H]
\centering
\includegraphics[width=1.0\textwidth]{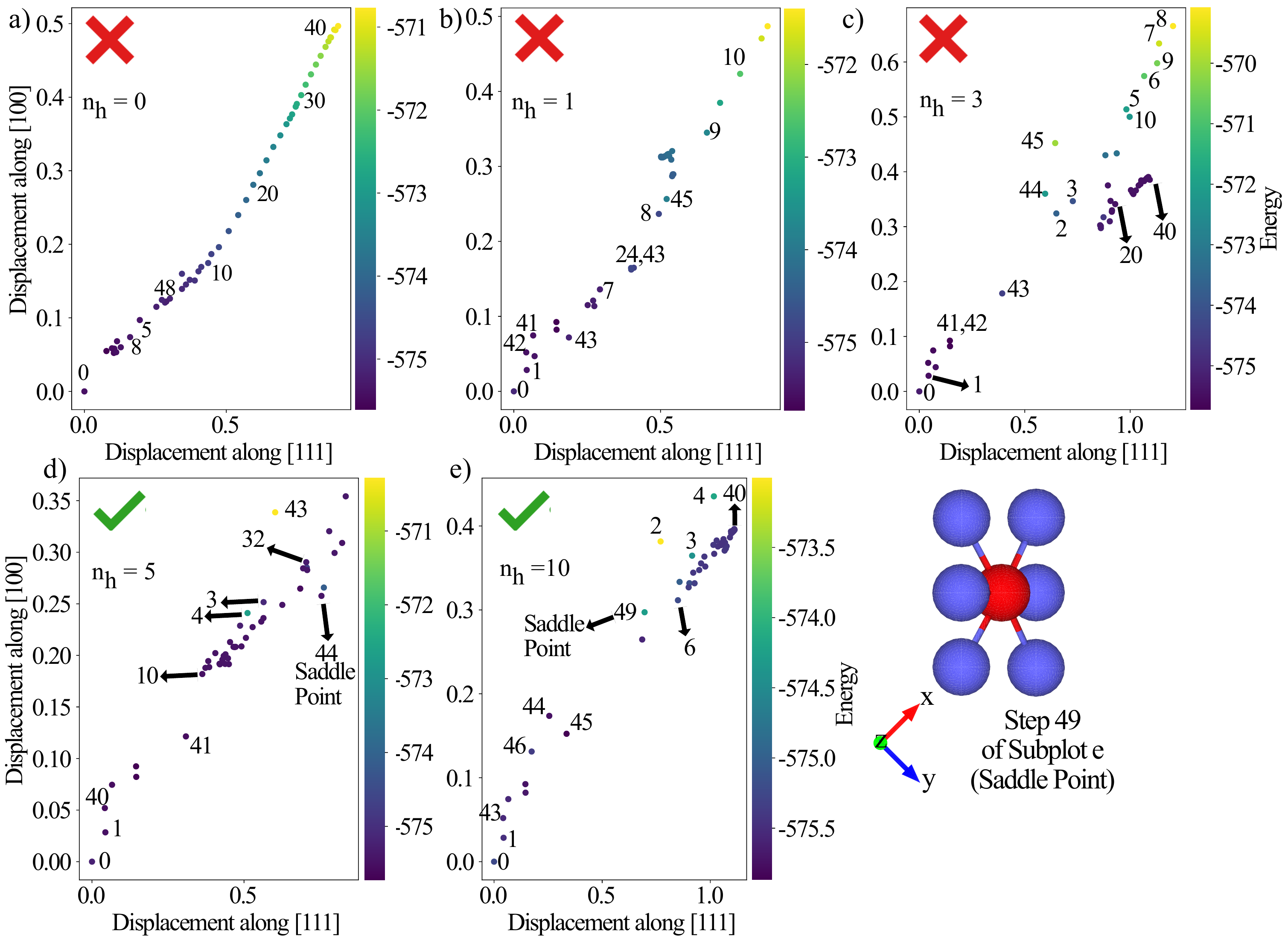}
\caption{The GPR-dimer trajectory projected along [100] and [111] during the saddle point search of self-diffusion in \textit{bcc} Mo with the translation hop frequency of a) zero, b) one, c) three, d) five, and e) ten. Color bars map the energy of different dimer steps.}
\label{fig:supp1}
\end{figure}


\section{The Inescapable Valley}
we explore the utilization of both the standard DFT-dimer and GPR-dimer methods for identifying the saddle point associated with the monovacancy diffusive jump in \textit{bcc} Mo. In contrast to the findings presented in the main article, our focus here is on starting the saddle point search from the atomic configuration located within the basin (or valley) on the energy surface. Specifically, this configuration corresponds to the \textit{bcc} supercell containing a vacant site. To perform the dimer search, we employ two implementations: the VTST code \cite{Henkelman2000climbing} and our in-house dimer implementation. In both implementations, a conjugate gradient (CG) scheme is utilized during the dimer search process.
\begin{figure}[H]
\centering
\includegraphics[width=.8\textwidth]{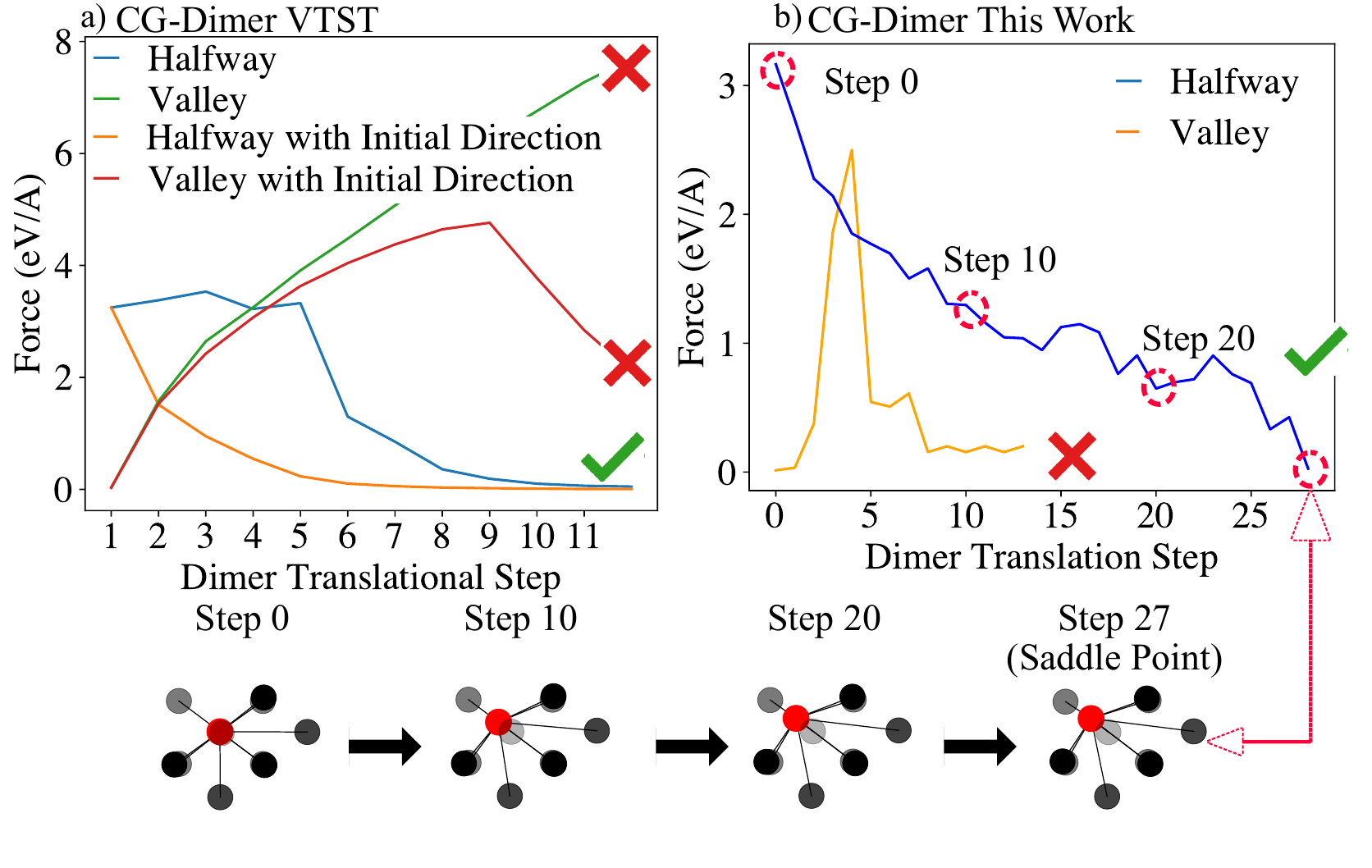}
\caption{DFT-dimer search for self-diffusion in \textit{bcc} Mo using a) the VTST code and b) our implementation of the dimer method. The dimer walks are launched, starting from an atomic configuration in the valley (or basin) of the energy surface or a geometrical interpolation of the atomic configuration halfway between the basin and the midway along the diffusive jump direction. In the VTST implementation, walkers are launched either with or without an initial guide direction for the translation step. Our implementation is exclusively run without a guide direction.}
\label{fig:dimer_valley}
\end{figure}
Figure~\ref{fig:dimer_valley}(a) illustrates the outcome of initiating the dimer from the basin configuration. In this case, the dimer fails to locate the saddle point, regardless of whether or not an initial direction is provided for its first translation step. Conversely, when the dimer is initiated from an atomic configuration interpolated a quarter away from the basin configuration, it successfully identifies the saddle point. This atomic configuration is shown as halfway in Figure~\ref{fig:dimer_valley} and corresponds to an state in 1/4th of the vector between initial and final states, in this case, along the [111] diffusive jump direction. Our in-house dimer implementation, as shown in figure~\ref{fig:dimer_valley}(b), also achieves the successful identification of the saddle point when starting from the halfway configuration without any guiding direction. However, if the dimer is initiated from the valley configuration, it is unable to successfully locate the saddle point and instead returns to the valley configuration. Further details regarding these observations are provided below.
\begin{figure}[H]
\centering
\includegraphics[width=1.0\textwidth]{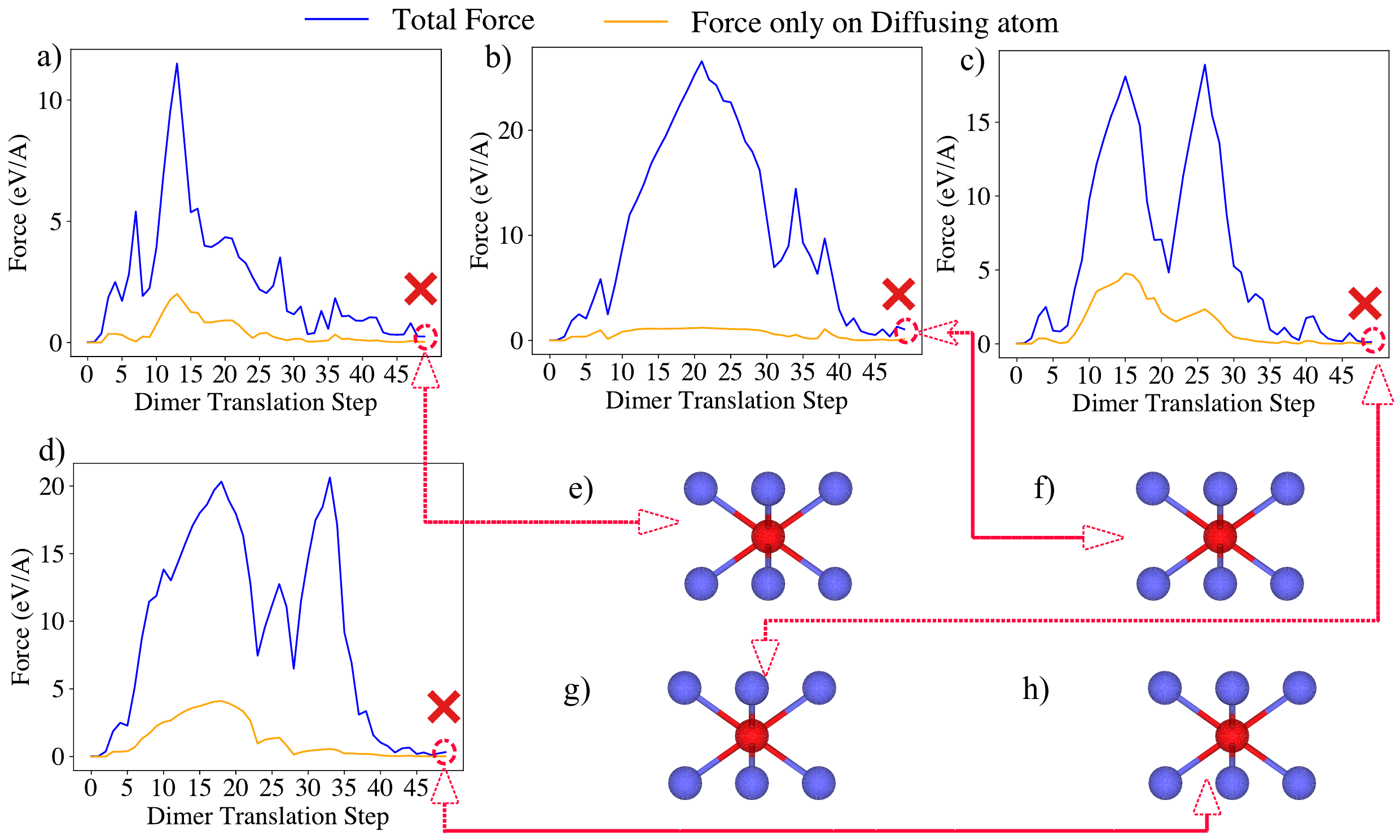}
\caption{GPR-dimer saddle point search for self-diffusion in \textit{bcc} Mo, initiated from the local minimum (or valley) configuration. The total force and diffusing atom force for a GPR-dimer with a) 1, b)3, c)5 and d) 10 translation hops. e-h) Atomic configuration in the vicinity of the diffusing atom (colored as red) at the end of the GPR-dimer search shown in a to d.}
\label{fig:GPR-valley}
\end{figure}

As anticipated, the GPR-dimer method produces comparable outcomes to the standard DFT-dimer. 
Figure \ref{fig:GPR-valley} presents the results of the GPR-dimer search initiated from the valley, employing an active region radius of $r_f=3\AA$ and initial training data consisting of the first three DFT-dimer translations ($n_i=3$). We investigate various translation hops, $n_h$, but in all cases, the GPR-dimer fails to locate the saddle point. As depicted in Figure \ref{fig:GPR-valley}(a-d), regardless of the translation hop frequencies, the GPR-dimer experiences substantial forces during the search, indicating exploration of high-energy regions on the energy surface away from the basin. However, in each case, the forces eventually diminish, suggesting that the GPR-dimer walker returns to the original atomic configuration in the valley (refer to Figure \ref{fig:GPR-valley}(e-h)).
%
\begin{figure}[H]
\centering
\includegraphics[width=1.0\textwidth]{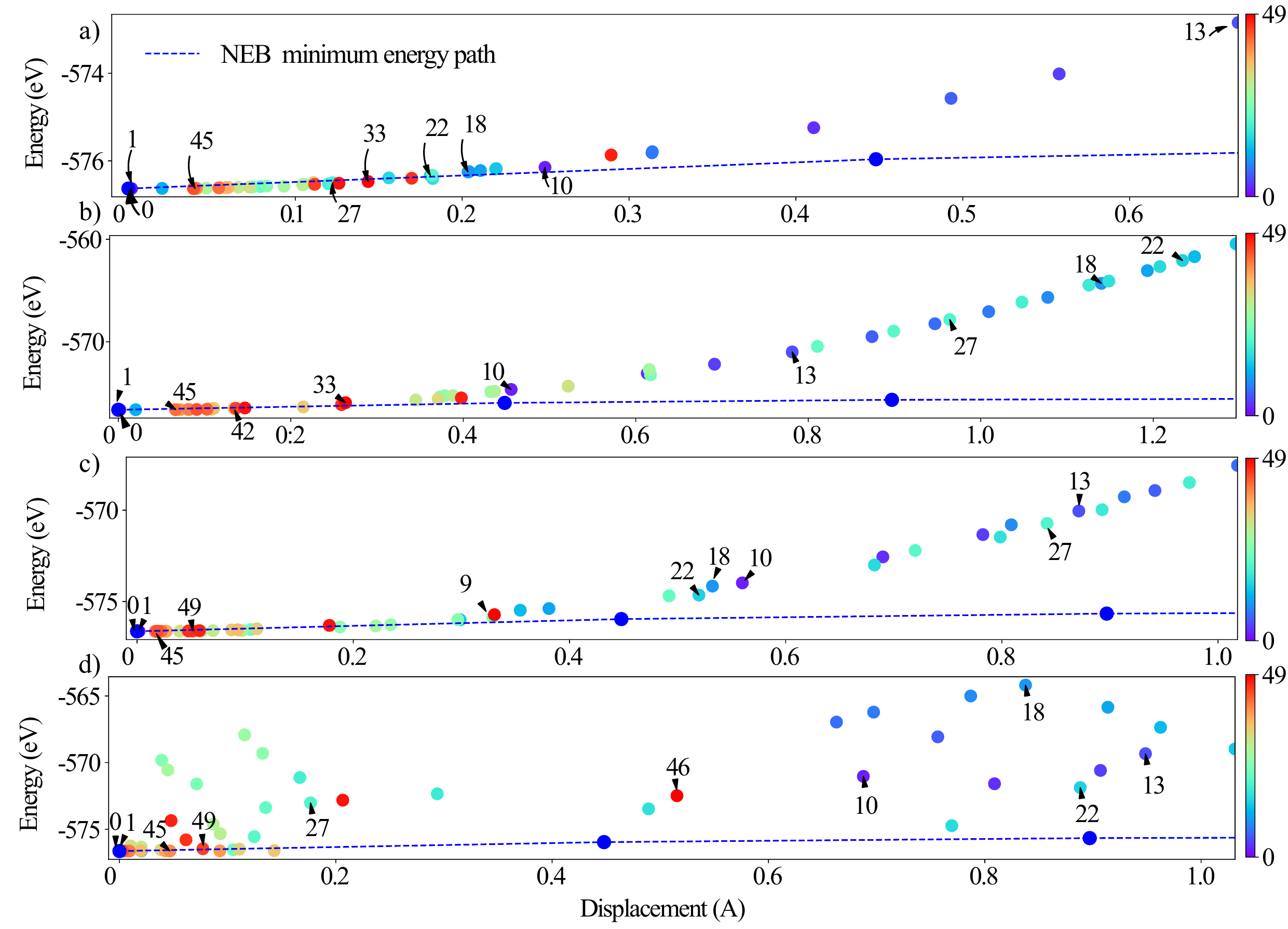}
\caption{Energy of the GPR-dimer versus its displacement along [111] direction for different numbers of translation hops. The GPR-dimer is initiated from the local minimum configuration for monovacancy diffusion in \textit{bcc} Mo.}
\label{fig-energyprofile}
\end{figure}
To provide further evidence that the GPR-dimer returns to the initial atomic configuration in the valley, Figure \ref{fig-energyprofile} illustrates the energy profile of the GPR-dimer plotted against its displacement projected along the [111] direction. Irrespective of the number of translation hops, the GPR-dimer walker explores high-energy regions on the energy surface but fails to locate the saddle point. Instead, it converges back to the initial atomic configuration. In all cases depicted in Figure \ref{fig-energyprofile}, the GPR-dimer is executed for 49 translation steps, excluding the GPR-guided hops in between. As observed in Figure \ref{fig-energyprofile}, the final few steps of all GPR-dimer walkers closely align with the original atomic configuration, as they are concentrated near zero displacement values.
\begin{figure}[H]
\centering
\includegraphics[width=1.0\textwidth]{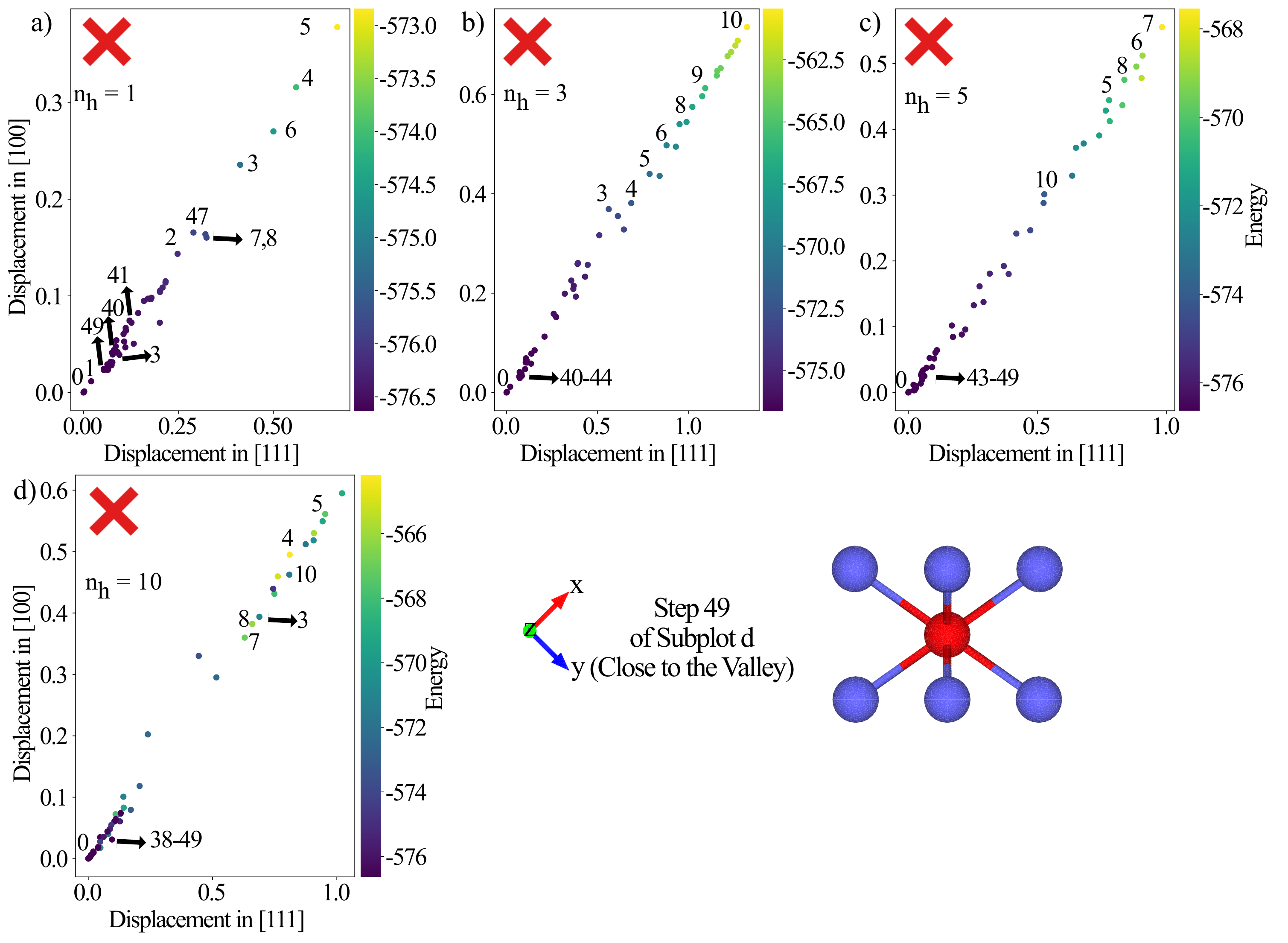}
\caption{The GPR-dimer trajectory projected along [100] and [111] during the saddle point search of self-diffusion in \textit{bcc} Mo. The GPR-dimer is initiated from the local minimum atomic configuration, which is associated with a vacancy residing on a \textit{bcc} site. Color bars map 
 the energy of different dimer steps.}
\label{fig:traj-second}
\end{figure}
Furthermore, in Figure \ref{fig:traj-second}, we present the trajectory of the GPR-dimer projected along [100] and [111] directions during the saddle point search. It is evident that regardless of the translation hops used, the walker consistently returns to the initial atomic configuration near the valley on the energy surface.

The obtained results indicate that both the DFT-guided dimer method and the GPR-dimer approach tend to get trapped in the valley. This limitation can be attributed to the inherent characteristics of the dimer algorithm, as discussed in the original dimer paper \cite{Henkelman1999dimer} and confirmed by our investigations. It should be noted that this limitation is not related to the GPR surrogate model itself. To overcome this challenge and explore the energy landscape more effectively, one can consider complementing the dimer search with biased search methods such as metadynamics \cite{Parrinello2002,Parrinello2011} or basin-hopping algorithms \cite{Doye1997}. These techniques can provide a more comprehensive exploration of the energy landscape and potentially overcome the trapping issue observed with the dimer method.
\newpage
\begin{table}[H]
    \centering
    \begin{tabular}{lll}
    \toprule
       & \textbf{DFT Calculations}  & \textbf{Computational Time (Core-Hour)}  \\
    \midrule
        GP-Dimer (1 hop)  & 48  & - \\
        GP-Dimer (3 hops)  & 45 & - \\
        GP-Dimer (5 hops)  & 44 & 17.89 \\
        GP-Dimer (10 hops)  & 49 & 19.2 \\
    \bottomrule
    \end{tabular}
    \caption{Effect of the number of translation hops on the computation effort of the GPR-dimer for \textit{bcc} Mo self diffusion ($r_f=3 \AA$).}
\end{table}

\begin{table}[H]
    \centering
    \begin{tabular}{lll}
    \toprule
       & \textbf{DFT Calculations}  & \textbf{Computational Time (Core-Hour)}  \\
    \midrule
        GP-Dimer (3\AA{})  & 49 & 19.2  \\
        GP-Dimer (5\AA{})  & 43 & 19.78  \\
        GP-Dimer (7\AA{})  & 43 & 22.36  \\
    \bottomrule
    \end{tabular}
    \caption{Effect of the active region radius on the computation effort of the GPR-dimer for \textit{bcc} Mo self diffusion ($n_h=10$).}
\end{table}

%